\documentclass[conference]{IEEEtran}

\usepackage{url}
\usepackage{cite}
\usepackage{xspace}
\usepackage{amsmath,amssymb,amsthm}
\usepackage[pdftex]{xcolor,graphicx}
\usepackage{adjustbox}
\usepackage{underscore}
\usepackage{authblk}

\newcommand{\djznt}[1]{}
\newcommand{\tmknt}[1]{}
\newcommand{\npdnt}[1]{}
\newcommand{\bhnt}[1]{}
\newcommand{\hxnt}[1]{}

\newcommand{\eg}{\textit{e.g.}\xspace,\xspace}

\newcommand{\TWIAD}{TWIAD\@\xspace}

\newcommand{\refsec}[1]{Section~\ref{#1}\@\xspace}
\newcommand{\reffig}[1]{Figure~\ref{#1}\@\xspace}
\newcommand{\reftab}[1]{Table~\ref{#1}\@\xspace}

\newcommand\Mark[1]{\textsuperscript#1}

\IEEEoverridecommandlockouts

\author[1,2]{Nolan Donoghue}
\author[1,2]{Bridger Hahn}
\author[1,2]{Helen Xu} 
\author[1]{Thomas Kroeger}
\author[3]{David Zage\Mark{@}}
\author[2]{Rob Johnson}
\affil[1]{Sandia National Laboratories\Mark{*}\thanks{\Mark{*}
Sandia National Laboratories is a multi-program
laboratory managed and operated by Sandia Corporation, a wholly
owned subsidiary of Lockheed Martin Corporation, for the
U.S. Department of Energy's National Nuclear Security Administration
under contract DE-AC04-94AL85000.

\Mark{@}Work performed while at Sandia National Laboratories.} Livermore, CA}
\affil[2]{Department of Computer Science, Stony Brook University, NY, USA}
\affil[3]{Intel Corporation, Santa Clara, CA, USA.}

\begin{document}

\title{\LARGE{Tracking Network Events with Write Optimized Data Structures}\\
\vspace{0.05in}
\large{The Design and Implementation of TWIAD: The Write-Optimized IP Address Database} 
}

\maketitle

\begin{abstract}
Access to network traffic records is an integral part of recognizing
and addressing network security breaches.  Even with the increasing
sophistication of network attacks, basic network events such as
connections between two IP addresses play an important role in any 
network defense.  Given the duration of current attacks, long-term 
data archival is critical but typically very little of the data is
ever accessed. Previous work has provided tools and
identified the need to trace connections. However, traditional
databases raise performance concerns as they are optimized for 
querying rather than ingestion.

The study of write-optimized data structures (WODS) is a new and
growing field that provides a novel approach to traditional storage
structures (\eg B-trees). WODS trade minor degradations
in query performance for significant gains in the ability to quickly
insert more data elements, typically on the order of 10 to 100 times
more inserts per second. These efficient, out-of-memory data
structures can play a critical role in enabling robust, long-term
tracking of network events.

In this paper, we present \emph{T}WIAD, the \emph{W}rite-optimized \emph{I}P 
\emph{A}ddress \emph{D}atabase.  TWIAD uses a write-optimized B-tree known as a 
B$^\varepsilon$ tree to track all IP address connections in a network traffic stream.
Our initial implementation focuses on utilizing lower cost hardware, demonstrating  
that basic long-term tracking can be done without advanced equipment. We tested TWIAD on a 
modest desktop system and showed a sustained ingestion rate of about 20,000 inserts per second.

\end{abstract}

\section{Introduction}\label{sec:intro}
The recent hack of the US government's Office of Personel Management (OPM)
exposed the personal information of millions of federal employees. The OPM breach
illustrates the critical role that network monitoring systems have and will
continue to play in cybersecurity. According to the Department of Homeland
Security (DHS), an intrusion-detection program called Einstein
was involved in the response to the breach~\cite{OPM}. 
DHS's Computer Emergency Readiness Team used the Einstein system to discover 
the recent hack at OPM. After OPM suffered a breach in March 2014, 
the agency beefed up its cybersecurity via a ``comprehensive network monitoring plan, 
through which OPM detected new malicious activity.''

Tracking network events continues to play an integral part of securing data and communication
networks.  This includes being able to store all of the connections occurring over a network, enabling 
applications such as intrusion detection and post-event forensics. Effective network event tracking 
requires data structures that can consistently provide strong ingestion performance guarantees over long periods of time.  Furthermore,
recent public compromises such as the OPM breach have highlighted the need for maintaining
records that cover years worth of traffic.

Many institutions use common data stores such as Hadoop~\cite{White:2009:HDG:1717298}
in server clusters on advanced hardware for network situational awareness.
Although this solution may be performant, it is costly because of the hardware
and maintenance involved. We wanted a lightweight solution with a smaller
start-up cost that maintained high performance guarantees.   

Other situational awareness solutions use large databases based on Bloom filters, 
a probabilistic data structure for approximate
membership queries \cite{bloom1970space}. Bloom filters are placed in front
of data structures such as B-trees and improve query performance by providing
constant-time, negative answers to point queries. Since our solution must be efficient 
for point and range queries, Bloom filters alone are insufficient because they
do not improve range query performance.  Moreover, Bloom filters
are only viable when they can be contained within RAM, limiting
their ability to keep long-term data that grows without bound.

Finally, it might seem that solid state drives (SSDs) solve many of the issues associated with 
network traffic ingestion since they have random I/O performance 
and latency orders of magnitude better than traditional spinning hard-disk
drives (HDDs). However, SSDs are substantially more expensive per gigabyte
than HDDs, which can be a prohibitive barrier to widespread use of SSDs for network
monitoring.

Furthermore, the issue of write amplification
occurs uniquely in SSDs because the actual physical size of the data written
to disk may be a multiple of the logical size of the data intended
to be written~\cite{hu2009write}. Flash memory uses relocate-on-write, 
which erases memory before it is rewritten
and requires garbage-collection because the erase operation is much less 
precise than the write operation. Therefore, writing to a SSD requires
accessing and moving user data and metadata more than once. Rewriting
on the disk reads, updates, and rewrites existing physical data to a new
location on the disk. Larger amounts of data must be erased and rewritten
than the actual amount required by the new data. Flash memory disk blocks
wear out as data is erased and rewritten, shortening the life of an SSD.
As a result of cost and write amplification, SSDs alone are not a solution
to the challenges of network monitoring on a large scale. Therefore, 
our solution must perform well on HDDs as well as SSDs. 

We therefore propose the use of write-optimized data structures (WODS)~\cite{brodal2003lower,
vitter2001external, brodal2010cache, graefe2004write, bender2007cache,
buchsbaum2000external} for I/O-efficient tracking of long-term network traffic.
WODS are designed to resolve the bottleneck caused by writing to disk and can ingest data up to two
orders of magnitude faster than traditional B-trees. 
Additionally, WODS mitigate write amplification in SSDs by writing
larger blocks of data to disk at a time and reducing the amount of physical
memory that needs to be deleted and rewritten.

Write-optimization has been successfully implemented in many commercial
key-value stores and databases~\cite{accumulo, hbase, leveldb, chang2008bigtable,
lakshman2010cassandra, tokudbmysql, tokumxmongodb}. Additionally, previous
research demonstrates the feasibility of write-optimization in indexing
core system components such as file systems~\cite{jannen2015betrfs}.

We present \emph{TWIAD}, a write-optimized database
tailored to IP address tracking. We use the mature B$^\varepsilon$-tree
implementation from Tokutek's Fractal Tree Index (ft-index)~\cite{tokudbmysql}
as the index under \TWIAD. While we initially focus on tracking IP addresses,
our system is generic and can easily be adapted to track other network data
such as domain names, complete URLs, and email addresses.  Our initial results show that a B$^\varepsilon$-tree-based index
provides a feasible, lightweight, and portable database for tracking
network events. Our contributions include:
\begin{itemize}
\item the application of write-optimized data structures to network
      event tracking
\item the design and implementation of a write-optimized IP address
      database (\TWIAD) and associated query tools 
\item an initial performance analysis of \TWIAD on basic hardware
showing a high ingest rate of 20,000 entries per second.
\end{itemize}

The rest of the paper is organized as follows: we review related work
and background information in \refsec{sec:background}, describe the
requirements and resulting design in in \refsec{sec:rnd}, present our results
in \refsec{sec:results}, discuss future applications of \TWIAD and
write-optimized data structures as a whole
in \refsec{sec:applications}, and conclude our work
in \refsec{sec:conclusion}.

\section{Background And Related Work}\label{sec:background}
In this section, we describe write-optimized data
structures, streaming databases, and existing network event tracking systems.

\subsection{Write-Optimized Data Structures}
Here, we cover write-optimized data structures
and their performance bounds. Specifically, we describe the B$^\varepsilon$-tree
and the reasons we have chosen it for network event tracking. The best WODS 
(including the B$^\varepsilon$-tree) subvert a trade-off between read and
write performance and instead can outperform B-trees.

The B-tree is a data structure where internal nodes 
have variable numbers of children within a predefined range. The elements
of the tree are maintained in sorted order at the leaves. Each internal
node has keys directing searches to the subtree associated with the query value.
B-trees support insertions, deletions, sequential accesses, and point queries
in $O(\log_BN)$ time.

\smallskip

\subsubsection{B$^\varepsilon$ trees}
A B$^\varepsilon$-tree is a B-tree with buffers
at each node. New insertions take place at the root buffer of the B$^\varepsilon$-tree. When
a node's buffer is filled, items are moved from that node's buffer to the buffer of
one of it's children---this process is referred to as flushing. The algorithms for point
and range queries are the same as those in a B-tree, but with a search through the
buffer of each internal node on a root-to-leaf path.

B$^\varepsilon$-trees have asymptotically better performance than 
B-trees. For example, consider a B-tree of $N$ elements where each node has
$B$ keys of constant size and where the the size of keys is far larger than
the size of the related data. Such a tree has fanout $B$ and therefore
has height $O(\log_BN)$. Therefore, inserts and searches will take $O(\log_BN)$ I/Os.
A range query with $k$ results then requires $O(\log_BN+\frac{k}{B})$ I/Os.

In contrast, a B$^\varepsilon$-tree has nodes of size $B$. Each internal node
of the tree has B$^\varepsilon$ children where $0$ \textless $\varepsilon \leq 1$. 
Each node has a ``pivot key'' for each child,
so the keys take up $B^\varepsilon$ space in each node. The remaining $B-B^\varepsilon$ space
in each node is used to buffer inserted elements. 

$\varepsilon$ is a tunable parameter that determines the tree's fanout. The tree's fanout is 
$B^\varepsilon$ and its height is $O(\log_{B^\varepsilon}N) = O(\frac{1}{\varepsilon}\log_BN)$.
Therefore, searches in a $B^\varepsilon$-tree are slower than those in a $B$-tree by a factor
of $\frac{1}{\varepsilon}$. However, whenever a node flushes elements to one of its children,
it moves at least $\frac{B-B^\varepsilon}{B^\varepsilon} \approx B^{1-\varepsilon}$ elements.
Each element must be flushed $O(\frac{1}{\varepsilon}\log_BN)$ (the height of the
tree) times to reach a leaf. Therefore, the amortized cost of inserting $N$ elements is 
$O(\frac{1}{\varepsilon B^{1-\varepsilon}}\log_BN)$. Furthermore, range queries cost 
$O(\frac{1}{\varepsilon}\log_BN + k/B)$ I/Os where $k$ is the number of elements 
returned by the query.

We present an example: consider $\varepsilon = 1/2$. Point and range query costs are now
$O(\log_BN)$ and $O(\log_BN + \frac{k}{B})$ respectively. Although these are the same asymptotically as
query bounds for B-trees, the insert cost for the $B^\varepsilon$-tree is 
$O(\frac{\log_BN}{\sqrt{B}})$, an improvement by a factor of $\sqrt{B}$ over
traditional B-trees.

Additionally, B$^\varepsilon$ trees have much larger nodes than B-trees. Larger nodes
improve range query performance because the data is spread over fewer nodes and therefore
requires fewer disk accesses to read it in. B-trees must use smaller nodes because every
new addition to the database requires that a node be completely rewritten. In
contrast, writes are batched in B$^\varepsilon$ trees, allowing their nodes to be much
larger than those of a B-tree --- for example, nodes in a B-tree are generally around 4
or 6KB, while a typical node in Tokutek's implementation of a $B^\varepsilon$-tree is 4MB.
As a result, the height of a B$^\varepsilon$ tree is not
much greater than that of a B-tree on the same data. Therefore, point query
performance in a B$^\varepsilon$ tree is comparable to point query performance in a B-tree.

For example, consider a key-value store of 1TB of data, with keys of size 128B and records
(key+value) of size 1KB --- assume that data is logged and that all updates in the log
are periodically applied to the main tree in batch. Assume that a common server has about
64GB of RAM.

First, we examine a B-tree with 4KB nodes given the above situation. The fanout of the
tree is 4KB/128B = 32. Even if all of the internal nodes of the tree can fit into RAM,
only a small fraction of the 1TB of leaf nodes can be held in cache. Given a sequence of
random insertions, most updates will require 2 I/Os --- 1 I/O to read in the target leaf
and another to write it back to disk.

In constrast, consider a B$^\varepsilon$-tree with branching factor 10 and nodes with size 1MB.
Again, all internal nodes can fit in cache, but the leaves must be held on disk. When items
are inserted into the tree, they are stored in the tree's root buffer. The root is cached, so
this action requires no I/Os. When an internal node becomes full and flushes to a non-leaf
child, the data structure requires two writes --- one to update the parent and one to
update the child. Since both nodes are cached, no reads are necessary. If an internal node
flushes its buffer to a leaf, one read is required to load the leaf into memory.
There will be 1TB/1MB=2$^{20}$ leaves. Furthermore, the tree has fanout 10, so its height
will be 1+$\log_{10}2^{20} \approx 7$. Therefore, each item is involved in 14 I/Os because it is
written and read once at each level of the tree.

While it may seem that this performance is worse than that of a B-tree, each flush in the
B$^\varepsilon$-tree moves $\sim$1MB/10$\approx$100kB of data, or around 100 items.
The data moved in each flush is approximately proportional to the node size divided
by the branching factor. Therefore, the amortized
cost of flushing an item to a leaf is 14/100. A B-tree requires 2 I/Os for each item, so
in our example the B$^\varepsilon$ tree can insert data 2/(14/100) $\approx$ 14 times faster 
than the equivalent B-tree. Furthermore, this speedup grows as key-value pairs get 
smaller as in connection log storage.

Both the B-tree and the B$^\varepsilon$ tree require a single I/O to read the corresponding
leaf in a point query. However, range queries can be much faster in a B$^\varepsilon$-tree
because the B$^\varepsilon$-tree seeks once for each leaf size. In our example, the B-tree
would need to seek every 4KB whereas the B$^\varepsilon$ tree would seek once every 1MB.

B$^\varepsilon$-trees can achieve further improved performance through upserts, an efficient
method for updating key-value pairs. If an application wants to update a value associated
with some key $k$ in the tree, it inserts a message $(k,(f, \Delta))$ into the tree, where
$f$ is some function that can be used to apply the change denoted by $\Delta$ to the old
value associated with $k$.

This message is inserted into the tree normally. However, when the message is
flushed from a node to one of its children $C$, the tree checks whether $C$'s buffer
contained the old value $v$ associated with the key $k$. If so, then the tree
replaces $v$ with $f(v, \Delta)$ and discards the upsert.

If the key $k$ is queried before the function from the upsert message is applied,
the B$^\varepsilon$-tree calculates $f(v,\Delta)$ while answering the query --- this does
not affect query performance because an upsert for a key $k$ will always be on the 
path from the root to the leaf containing $k$. Therefore, upserts can improve update speed by orders 
of magnitude without reducing query performance.

\subsubsection{Log-structured Merge Trees} The log-structured merge tree (LSM tree)~\cite{o1996log,sears2012blsm} 
is another write-optimized data structure. There are
many variations on the LSM tree, but generally they have a logarithmic number of
indices (data structures, e.g., B-trees) of exponentially increasing size. Once an index at one level
fills up, it is flushed and merged into the index at the next largest level. Commercial 
write-optimized databases often use LSM trees.

Although LSM trees can have the same asymptotic complexity as a B$^\varepsilon$-tree, queries in
a na\"{\i}ve implementation of a LSM tree can be slow, as shown in Table~\ref{table:wods}. Steps have been taken to
improve the query performance of LSM trees---of note, many implentations of LSM trees now
use Bloom filters~\cite{bloom1970space} at each index~\cite{accumulo, hbase, leveldb, chang2008bigtable,
lakshman2010cassandra}. Point queries in LSMs with Bloom filters have been reported to improve to
$O(\log_BN)$, therefore matching B-tree point query performance.

However, Bloom filters do not help with range queries, because the successor of any key may be
at any level of the data structure. Furthermore, the viability of Bloom filters degrades with
upserts. To compute the result of a query, all relevant upserts must be applied to the key-value
pair. If there are many possible upserts at each level of the LSM tree, searches need to
be performanced at each of those levels. LSMs only match B-tree query performance in specific cases, while
B$^\varepsilon$-trees match B-tree query times in general.

Since range queries are common in network event detection systems, we chose
B$^\varepsilon$-trees as the underlying data structure for TWIAD because of
LSM range query performance.

\begin{table}
\centering
\caption{Asymptotic I/O costs of various write-optimized data structures}
\label{table:wods} 
\resizebox{\columnwidth}{!} {
\begin{tabular}{ c c c c c }
  \hline	
  Data Structure & Insert & \parbox[t]{1.5cm}{Point Query\\w/o Upserts} & \parbox[t]{1.5cm}{Point Query\\w/ Upserts} & Range Query \\
  \hline	
  \hline
  B-tree & $\log_BN$ & $\log_BN$ & $\log_BN$ & $\log_BN+\frac{k}{B}$ \\ [2ex]
  LSM & $\frac{\log_BN}{\varepsilon B^{1-\varepsilon}}$ & $\frac{\log^2_BN}{\varepsilon}$ & $\frac{\log^2_BN}{\varepsilon}$ & $\frac{\log^2_BN}{\varepsilon}+\frac{k}{B}$ \\[2ex]
  LSM+BF & $\frac{\log_BN}{\varepsilon B^{1-\varepsilon}}$ & $\log^2_BN$ & $\frac{\log^2_BN}{\varepsilon}$ & $\frac{\log^2_BN}{\varepsilon}+\frac{k}{B}$ \\[2ex]
  B$^\varepsilon$-tree & $\frac{\log_BN}{\varepsilon B^{1-\varepsilon}}$ & $\frac{\log_BN}{\varepsilon}$ & $\frac{\log_BN}{\varepsilon}$ & $\frac{\log_BN}{\varepsilon}+\frac{k}{B}$ \\[2ex]
  \hline  
\end{tabular}
}
\smallskip

\end{table}

\subsection{Streaming Databases}

A great deal of progress has also been made in the related field of stream processing engines (SPE).
While data stream managers have important applications in network event tracking, most
of the literature is focused on developing query schemes and algorithms~\cite{abadi2005design, 
babu2001continuous, golab2003issues,carney2002monitoring}. Some of the issues identified by
the streaming community such as approximate query results, updating query results over time,
and dynamic query modification are outside of the scope of this paper.

Our vision for \TWIAD is that of a write-optimized streaming database for network event tracking.
Connection logs are constantly fed into the database --- the main goal is to process
a large volume of data consistently over time while maintaining ingestion guarantees. 
Streaming research has important applications to network event tracking and write-optimization.
We are currently focusing on optimizing ingestion and leave the integration of
results from streaming research as future work to optimize queries.

\subsection{Network Event Tracking Databases}

Previous development has also been done on large databases for network traffic monitoring.
Network traffic monitoring solutions differ from traditional relational
database management systems (RDBMSs) in the following ways:
\begin{enumerate}
 \item The data and storage must be stream-oriented. Fast ingestion and sequential
 access are important, while fast random access and concurrency control are not.
 \item Since network traffic data is usually only used a few times (or even once), load
 time is a significant cost. Therefore, the database must maintain data integrity 
 over long periods of time while still loading streams of data into the database.
 \item Network connection logs are aggregations of many small records with fields a 
 few bytes wide, so per-tuple overhead in RDBMSs can lead to a prohibitive cost in space.
\end{enumerate}

Prior work on network event tracking databases has focused on developing query languages
for streams. Systems like Gigascope~\cite{cranor2003gigascope} and Tribeca~\cite {sullivan1998system}
propose query languages for complex analysis. The inventors of the existing systems note that
performance for a stream database is measured by how high the input stream(s) rate can
be before it begins dropping data, not how fast the database can answer queries. Specifically,
Cranor et al. observe that ``touching disk kills performance---not because it is slow
but because it generates long and unpredictable delays throughout the
system.'' Our system,
\TWIAD is designed to specialize in ingesting data quickly and predictably.

\subsection{Write-Optimized Intrusion Detection Systems}

Some IDS companies are beginning to offer services that include
write-optimized databases. For example, Countertack, an IDS software
company, uses big data analytics from Cloudera, which is built on
Hadoop and HBase~\cite{countertack}. Similarly, Google’s Stenographer 
does simple packet capture and uses LevelDB for storage~\cite{steno}. It is
designed to write packets to disk quickly and not well suited to reading
back large amounts of packets. Finally, Hogzilla is another open source
IDS supported by Snort, Apache Spark, and HBase\cite{hogzilla}.

While these systems are important steps towards using write optimization in
network event tracking, we wanted to build a lighter-weight and simple tool
that processes logs while still leaving the user freedom to determine what kind
of analytics they want to do on the data.

\section{Requirements and Design}\label{sec:rnd}
TWIAD is a write-optimized database built on Tokutek's Fractal Tree
index, an implementation of a B$^\varepsilon$-tree. 

\subsection{Requirements}

We designed TWIAD to leverage the performance strengths of
B$^\varepsilon$-trees.  We focused primarily on ingestion
performance---the database will answer queries, but most of the
computation is spent on inserting events documented in connection logs
into the database.

\subsection{Software}

We used the publicly available B$^\varepsilon$-tree implementation (ft-index) from 
Tokutek~\cite{ftindex}.

Our contribution is a system designed to mediate between network connection
data and the underlying B$^\varepsilon$-tree while simultaneously answering queries.
We chose to implement this layer in C because the original index was in C and
we wanted to maximize performance without going through other languages.

One of our design goals for the database is for the index to be ``loosely coupled'' with the IDS.
Therefore, we chose to implement it as a tool for processing logs from the IDS 
rather than integrating it with the IDS. That also means that the IDS can handle bursts, 
since it is always just logging. The indexer can run in the background at low priority, 
it can fall behind and then later catch up when the system is less busy.

\subsection{Database Design}

The database is designed specifically to store network connection logs
--- we used logs from the Bro IDS system, a sessionization and
intrusion detection tool.  Fields from a typical connection log and an
example row are provided in \reftab{table:bro}.  The database can
easily be extended to ingest other types of logs and
inputs.

We want to access any connection information regarding an IP in a query,
regardless of whether it was the origin or responder IP. Therefore,
we insert every row in the connection logs twice --- once with the 
the origin and responder IP address and port in the order of the original row as
specified in \reftab{table:key} and again with the origin and responder
IP addresses and ports switching order in the key as detailed in \reftab{table:rkey}.
The value is the remaining fields not included in the key. 

As a result, two database entries are created for each row in a connection log.
To indicate which of the entries is ``reversed'' (i.e. the order of the origin and
responder ports and addresses was switched), we set the byte isReversed to 1 if
the entry was reversed and 0 otherwise.

The key and value formats are described in \reftab{table:key}
and \reftab{table:value} are based on Bro log format. 
We chose this format in order to query
specific IP addresses as well as range over time for that IP
address. Furthermore, we appended the other fields to the key 
to prevent the possibility of collisions.

\begin{table}
\centering
\caption{Example Row Structure of Bro Connection Log}
\label{table:bro} 
\resizebox{\columnwidth}{!} {

\begin{tabular} {c | c c c c}
field & ts & uid & id.orig_h & id.orig_p \\
type & time & string & addr & port \\
\hline
value & 980997832.690939 & Ch80yl33lQkOymHcab & 209.11.146.100 & 111 \\
\hline
\hline
field & id.resp_h & id.resp_p & proto & service\\
type & addr & port & enum & string \\
\hline
value & 0.254.205.104 & 36831 & tcp & - \\
\hline 
\hline
field & duration & orig_bytes & resp_bytes & conn_state \\
type & interval& count & count & string\\
\hline 
value & 0.075242 & 0 & 0 & RSTOS0 \\
\hline
\hline
field & local_orig & missed_bytes & history & orig_pkts\\
type & bool & count & string & count \\
\hline
value & - & 0 & HR & 2 \\
\hline
\hline
field & orig_ip_bytes & resp_pkts & resp_ip_bytes & tunnel_parents \\
type & count & count & count & set[string]\\
\hline
value & 84 & 0 & 0 & (empty) \\
\end{tabular}
}
\end{table}

\begin{table}
\centering
\caption{Key Design}
\label{table:key} 
\resizebox{\columnwidth}{!} {
\begin{tabular} {c | c | c}
Byte Range & Length & Field \\
\hline
{[}0, 3{]} & 4 & Origin IP \\
{[}4, 11{]} & 8 & Timestamp (*10000 second since epoch) \\
{[}12, 14{]} & 3 & Origin Port \\
{[}15, 18{]}& 4 & Destination IP \\
{[}19, 21{]} & 3 & Destination Port \\
\end{tabular}
}
\end{table}

\begin{table}
\centering
\caption{Value Design}
\label{table:value} 
\resizebox{\columnwidth}{!} {

\begin{tabular} {c | c | c}
Byte Range & Length & Field \\
\hline
{[}0, 3{]} & 4 & Protocol \\
{[}4, 11{]} & 8 & Duration \\
{[}12, 19{]} & 8 & Origin Bytes \\
{[}20, 27{]} & 8 & Response Bytes \\
{[}28, 32{]} & 5 & Connection State \\
{[}33, 36{]} & 4 & Origin Packets Bytes (sans header) \\
{[}37, 40{]} & 4 & Response Packets Bytes (sans header) \\
{[}41{]} & 1 & isReversed \\
\end{tabular}
}
\end{table}

\begin{table}
\centering
\caption{Reversed Key Design}
\label{table:rkey} 
\resizebox{\columnwidth}{!} {

\begin{tabular} {c | c | c}
Byte Range & Length & Field \\
\hline
{[}0, 3{]} & 4 & Destination IP \\
{[}4, 11{]} & 8 & Timestamp (*10000 seconds since epoch) \\
{[}12, 14{]} & 3 & Destination port \\
{[}15, 18{]} & 4 & Origin IP \\
{[}19, 21{]} & 3 & Origin Port \\
\end{tabular}
}
\end{table}

\subsection{Client Architecture}

We designed a client-side CLI tool to relay queries from
an authorized user to the server and to receive and display
results. The goals of the CLI tool were to 
easily integrate into users' workflows and respond to queries.
Results of the queries are given in some value-separated format for easy
user processing.

Common queries include searching for connections with a specific IP
address or subnet or over some time range.

We provide a few examples of usage:

\noindent\texttt{/twiad:twiad-client --ip 18.281.23.9}

The first query requests all connections involving a
specific IP address over all time stored
in the database.

\noindent\texttt{/twiad:twiad-client --subnet 13.48.133.201/8 --year}

The second query requests all connections involving the specified
subnet over the past year. There are flags available to query
the past week, month, quarter, and year.

\noindent\texttt{/twiad:twiad-client --subnet 13.48.133.201/16 --start 2010:1:2:10:30:5}

The third query requests all connections involving the specified
subnet starting from the beginning of the time period after the start
flag. Times are entered in the format yyyy:mm:dd:hh:mm:ss. For example,
the above time is January 2 10:30:05 0700 MST 2010.

The tool also allows the user to specify the end of a requested time period
with \texttt{--end}.

\section{Results}\label{sec:results}



To benchmark TWIAD we used the publicly available backscatter dataset
from the Center for Applied Internet Data Analysis
(CAIDA)~\cite{backscatter}. We ran tests on several basic PCs and
measured the number of inserts completed every minute as well as
tracking the duration of insert transactions.  While our systems
initially saw high insertion rates as the size of our database grew we
stabilize to a steady state of approximately 20,000 inserts per second.
We benchmarked both the fractal tree index ($B^\varepsilon$-tree) and
Berkeley DB ($B$-tree). 

To provide a test data set, we generated Bro connection logs from the
backscatter dataset. The dataset consists of collections of responses
to spoofed traffic sent by denial-of-service attack victims and
received by the UC San Diego (UCSD) Network Telescope. Data was
collected between 2001 and 2008. While not perfectly representative
of normal network traffic it certainly represents actual network
traffic and presents a challenging data set. 

We ran our tests on a basic desktop PC with 32 gigabytes of RAM and
Intel i7-2600 CPU at 3.40GHz.  This system is representative of a typical
older desktop that an organization could repurpose as an IP address database system.

\subsection{Ingestion Performance}

As a basic metric to monitor our system’s ingestion rate, we recorded the amount 
of time it takes to insert a constant number of rows (e.g., 100,000). That is, 
we keep track of the time that it takes to insert {\em each} 100,000 rows.

\begin{figure}
  \centering
  \caption{Insertion rate vs. number of inserts}
\includegraphics[width=\columnwidth]{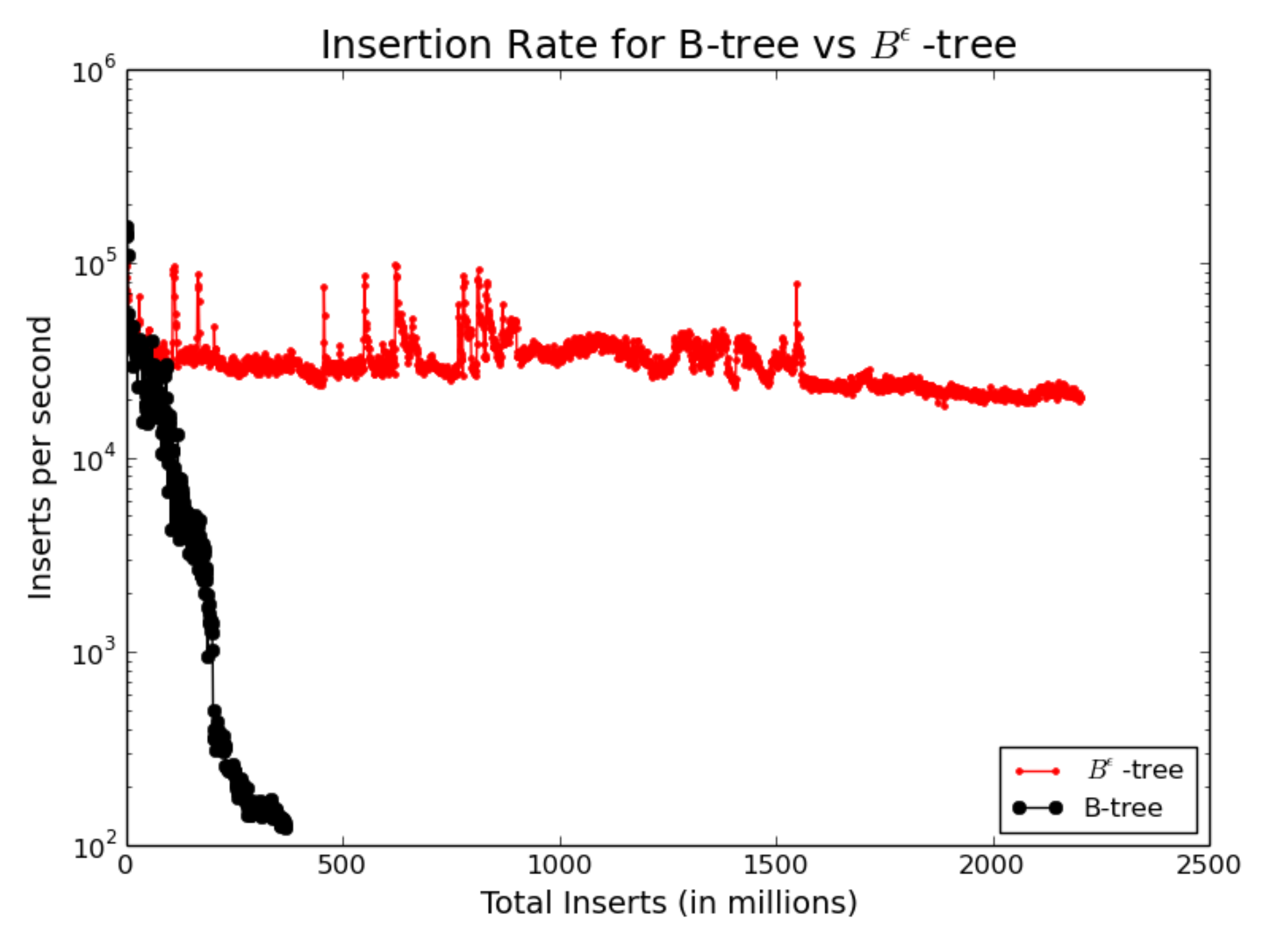}
  \label{fig:insertionRate} 
\end{figure}

From this we calculate the average number of inserts
per second over that period. Figure~\ref{fig:insertionRate} shows the
average insertion rate plotted against total number of database entries inserted
for both the fractal tree and traditional $B$ tree.

We can see that the database built on the fractal tree index (B$^\varepsilon$-tree)has a sustained 
insertion rate of 20,000 rows per second for over 1 billion rows. in contrast, the ingestion
rate for the database built on BerkeleyDB (B-tree) is similar to that of the fractal
tree index in the beginning but quickly and severely drops off to about 100 inserts per second.

\subsection{Query Response Time}

\begin{figure}
  \centering
  \caption{Query Performance Measurements (121.5 M entries in DB)}
\includegraphics[width=\columnwidth]{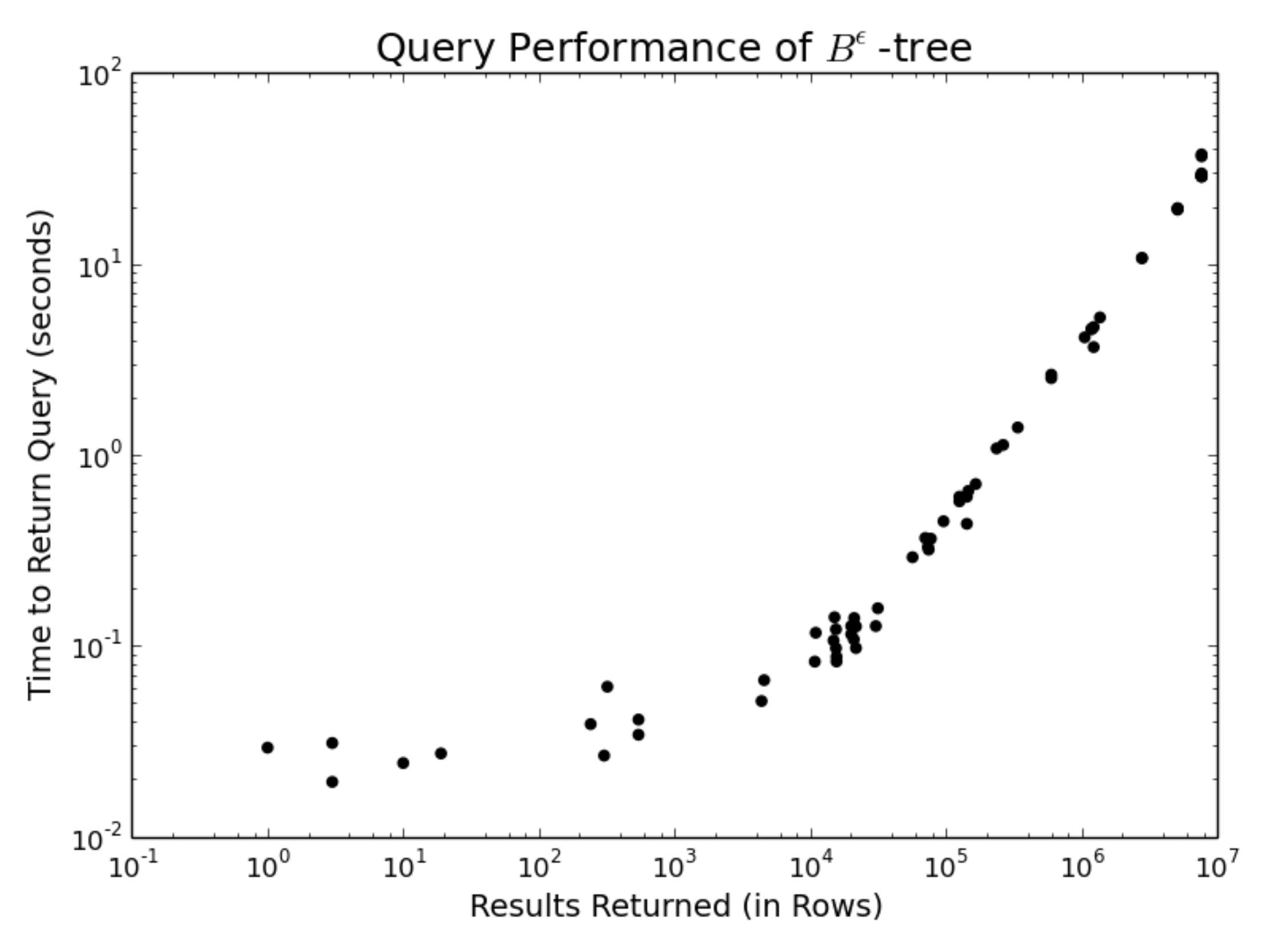}
  \label{fig:queryResp} 
\end{figure}

We tested a few queries around the size range that a typical user might execute. In our experience,
a simple point query came back in well under one second on a database with around 121.5 million entries.
Figure~\ref{fig:queryResp} shows the
time that it took a database using a fractal tree index with about 121.5 million
entries to answer range and point queries. 
We observe that reasonably sized queries that network analysts generally
expect return in well under a second. Queries are fast because write-optimization makes indexing
more efficient. 

As shown in table~\ref{table:querycomp}, queries for IP addresses in our database
generally return in under a second, while the same search using grep takes multiple seconds
or even minutes. Query times on the order of tenths of a second will allow network analysts to more quickly discover
and respond to network events. This is a significant improvement over a common strategy of searching through connection
logs.

\begin{table}[h]
\begin{center}
\caption{Fractal Tree Query vs Grep Performance (121.5 M Database Entries)}
\label{table:querycomp} 
\begin{tabular} {c | c | c | c}
IP Queried & Rows Returned & Time (grep, s) & Time (twiad, s)\\
\hline
0.183.158.39 & 3 & 6.242 & 0.030878 \\
202.178.243.254 & 7697311 & 796.276 & 37.528948 \\
61.132.23.66 & 600328 & 160.858 & 2.630128 \\
210.117.64.88 & 15513 & 10.386 & 0.096910 \\
210.117.64.222 & 20106 & 10.833 & 0.114642 \\
210.117.64.16 & 21005 & 35.084 & 0.126801 \\
210.117.64.84 & 21785 & 11.350 & 0.126013 \\
208.149.123.22 & 543 & 7.605 & 0.040972 \\
195.158.245.52 & 126186 & 16.253 & 0.570452 \\
200.42.64.226 & 15661 & 7.372 & 0.082603 \\

\end{tabular}
\end{center}
\end{table}

\section{Applications}\label{sec:applications}
\subsection{Intrusion Detection}

A great deal of progress has been made on network intrusion detection
systems that monitor network traffic for predefined patterns and
alert system administrators when potentially problematic network
traffic is detected. Previous work focuses on machine learning algorithms
and frameworks for intrusion detection~\cite{mukherjee1994network,
lee1998data, bass2000intrusion}. Additionally, systems like Bro
and Snort~\cite{bro, roesch1999snort} are lightweight intrusion-detection 
tools that take steps toward resolving issues with complex deployment and high cost. 

TWIAD has important applications in network intrusion detection. It allows
queries over long time periods and can efficiently store large amounts of
data while maintaining insertion performance guarantees. Users can ask for
all instances of an IP or subnet over a time range and analyze the connections
related to a potentially suspicious or adversarial IP. Efficiently storing
and having access to previous connection logs allows network analysis to
examine threats and take steps towards recovery.

\subsection{Data Visualization}

Write-optimized databases for network event tracking can also be used in
data visualization efforts on network traffic. Previous work with 
visualization focuses on the graphical representation of network flows~\cite{krasser2005real}.
Combining previous work on pattern detection and visualization with
faster packet storage will improve the scope and utility of the representations.

\section{Future Work}\label{sec:futurework}
While \TWIAD has seen success in processing long-term large-scale 
network traffic events, there is still much to be done.

For example, we would like to include applications of write-optimization
to sessionization tools such as Bro. The power of upserts would be especially
beneficial because the row corresponding to a connection could be efficiently
updated as packets arrive. Although our implementation did not take advantage of upserts 
because we did not need to update rows, we can leverage the performance gains resulting from upserts
in future work in updating the database or deleting from it.

In order to learn more about our system's performance, 
we would like to continue testing against other large-scale database
systems such as Hadoop, Accumulo, and Cassandra to determine the relative
performance of \TWIAD against currently popular solutions for analysis
of large datasets. Future experiments will also include tests on more robust hardware to
determine the performance of write-optimized data structures
on a variety of test systems and where they provide the greatest
speedup.

As discussed in \refsec{sec:background}, previous related work explores
the development of query languages for network event tracking. Although our current
work focuses on ingestion performance, we will use other results to
improve query performance and allow more specific requests to the database.
Thus, we leave a comparison of \TWIAD against other network event and stream tracking databases
such as Gigascope, Borealis, and Tribeca as future work.

Additionally, we will perform experiments to analyze possible adversarial
attacks on the data structure. The amortized cost of inserting new elements
in the database is at least an order of magnitude faster than traditional
structures such as B-trees. As the B$^\varepsilon$-tree grows large, the
time to flush down a root-to-leaf path grows because the height of the tree
increases. The flushing mechanism may trigger a cascade effect that requires
multiple flushes lower in the tree when the root node is flushed. Future
experiments will analyze data loss on a stream of network connections
over time. We will explore whether there is a data pattern or speed of
arrival that triggers a catastrophic sequence of buffer flushes and I/Os
that impact database performance. For example, the database will experience
a burst in connections during a denial-of-service attack and has to keep
up by ingesting the connection data.

\section{Conclusions}\label{sec:conclusion}
TWIAD demonstrates the potential of write-optimized data
structures for network event tracking. We have created a 
lightweight, portable system that can be installed on a simple
server and monitor network traffic. Furthermore, we were able to 
track connections with a generic server and modest hardware.
For smaller institutions with moderate resources, \TWIAD can easily
be installed on a spare machine, store information about connections,
and maintain situational awareness of the network. We have seen
sustained ingestion rates of around 20,000 inserts per second that
support the feasibility of \TWIAD as a real-time network event tracking
system.

Future work will focus on additional testing against other
systems and improvements to make the database more amenable
to data analysis. Our work indicates that future applications
of write-optimization in network security are likely to
improve performance.

\section*{Acknowledgments}
\addcontentsline{toc}{section}{Acknowledgments}\label{sec:acks}
We thank the engineers at Tokutek for developing and
open-sourcing their B$^\varepsilon$-tree implementation that
TWIAD is built on. Furthermore, we would like to thank Cindy Phillips, 
Jonathan Berry, Michael Bender, and Prashant Pandey for their
insights and thoughtful discussions. This work was supported by the 
Laboratory Directed Research and Development Program at Sandia National Laboratories.

\bibliographystyle{IEEEtran}
\bibliography{../bib/badgers-15}

\end{document}